\documentclass[showpacs,prd,amsmath,amssymb]{revtex4}
\usepackage{graphicx}
\usepackage{epsfig}
\usepackage{graphics}
\usepackage{amsmath}
\usepackage{dcolumn}
\usepackage{bm}
\usepackage{algorithmicx}

\begin{document}

\title{Effective Field Theory description of Phantom Fields}


\author{Elcio Abdalla}
\email{eabdalla@usp.br}
\affiliation{Instituto de Fisica, Universidade de S\~ao Paulo}
\begin{abstract}
In this work I show that a simple Field Theory on a non trivial gauge
background may behave as a phantom field and contribute to an effective $w<-1$
state equation fluid contribution to cosmology.
\end{abstract}

\maketitle

Cosmology is today a rapidly evolving field of research, with an enourmous number of ideas and highly unexpected results to be explained. Indeed, the description of an unexpected Dark Sector, especially Dark Energy, represents a huge step towards a new physics, starting from the conservative point of view of an extension of the Standard Model of particle physics with a positive Cosmological Constant \cite{1}, or also with new developments in relation to the Standard model, considering the existence of new fields (not earlier foreseen) and interactions thereof to describe a full fledged Dark Sector \cite{WangAbdallaAttrioP}, or even a completely new formulation of gravity for cosmology.

In general one expects Dark Energy to be described by a cosmological constant (as in the Standard Cosmological Model) or by a Scalar Field (as in quintessence) \cite{WangAbdallaAttrioP}. There are other descriptions, but generally speaking one takes for granted that, as in the case of a scalar field with self interaction the phantom behaviour is avoided, as one sees from a scalar field $\varphi$ with a self interacting potential $V(\varphi)$ (here independent of position)
\begin{equation}
\left.\begin{matrix} 
\rho =& \frac 12 \dot\varphi^2 + V(\varphi) \\
p=& \frac 12 \dot\varphi^2 - V(\varphi )  
\end{matrix}
\right\rbrace \qquad {\hbox {thus}} \quad w \equiv \frac p{\rho} > -1\quad .
\end{equation}
Phantom fields  are thus only expected in the presence of negative norm objects, what we try to avoid but may be important to describe elements of current phenomenology\cite{samisinghdadhich2003}\cite{wanggongabdalla2005}. The fact that the phenomenology of the equation of state of Dark Energy to favour, in some cases, $w<-1$ is a strong ndication of the presence of phantom fields.

However, a negative equation of state, can rather indicate instabilities generated by background fields. This is the point we wish to stress in this letter with some examples which may be physically achieved.

The exact value of $w$ or its derivative with respect to the cosmological time or space are largely unknown \cite{PlanckandotherPapers}. Although all observations are compatible with $w=-1$, the error bars are large enough to allow several parametrizations with respect to the redshift ($z$), not to speak about space dependence, which is completely unknown in the case where Dark Energy can clump. For models with scalar fields playing the role of Dark Energy the question is not answered \cite{LucasAndreAbdalla}.

A scalar field with $w=\frac p{\rho},-1$ can be constructed for unstable field theory configurations, such as a negative mass term, that appear  in Higgs models. Indeed, for an effective negative mass term $m^2 = -m_ {eff}^2$ in a scalar Lagrangian as in
\begin{equation}
{\cal L}= -\frac 12 \partial^\mu\varphi\partial_\mu\varphi +\frac 12 m_{eff}^2\varphi^2\quad ,
\end{equation}
we have, for a Minkowski metric,
\begin{equation}
w=\frac{\frac 12 \dot\varphi^2-\frac 12 \nabla\varphi^2 +\frac 12 m_{eff}^2\varphi^2}{\frac 12 \dot\varphi^2+\frac 12 \nabla\varphi^2 -\frac 12 m_{eff}^2\varphi^2}\quad .
\end{equation}
Therefore, there may be configurations, with small values for the derivatives, where $w<-1$.

There are also configurations where an effective quadratic term in the scalars can be generated from background fields. Indeed, let us consider a complex scalar interacting with a gauge field,
\begin{equation}
{\cal L}= -\frac 14 F_{\mu\nu} F^{\mu\nu} + \overline{D^\mu\phi}D_\mu\phi - V(\phi)\quad .
\end{equation}
We consider a background field expansion
\begin{eqnarray}
A_\mu&=&A_\mu^{(0)} + a_\mu\quad ,\nonumber\\
\phi &=&\phi^{(0)}+ \varphi\quad .
\end{eqnarray}
The Lagrangian expansion up to second order is
\begin{eqnarray}
{\cal L} &=& {\cal L}^{(0)}(A_\mu^{(0)},\phi^{(0)}) -\frac 14 F_{\mu\nu}(a_\mu)^2
\nonumber\\
&+& e^2a_\mu^2 \overline \phi^{(0)}  \phi^{(0)} + iea^\mu\left(\overline \phi^{(0)} D_\mu\varphi -\overline{D_\mu \phi^{(0)}\varphi}+\overline\varphi
D_\mu \phi^{(0)} -\overline {D_\mu\varphi }\phi^{(0)}\right)\nonumber\\
&+& \left( -\frac 12 V_0^{\prime\prime}\varphi^2-\frac 12 \overline \varphi^2 V_0^{\prime\prime}\overline\varphi^2 - V_0^{\prime\prime}\overline\varphi \varphi  
\right)\nonumber\\
&+& \overline{D_\mu^{(0)}\varphi} D^{\mu {(0)}}\varphi\quad .
\end{eqnarray}
The scalar field quadratic term, in the presence of the background fields is thus
\begin{equation}
\int d^3x\left\lbrack m^2 -e ^2 A_\mu^{(0)}A^{\mu (0)}  \right\rbrack\overline\varphi\varphi\quad .
\end{equation}
This term may lead to instabilities and $w < -1 $ effective solutions. Indeed, consider the Nielsen-Olesen solution \cite{NielsenOlesen}
\begin{equation}
\vec A^{(0)}(\vec x) = \frac{\vec x\wedge \hat e_3}{r}A^{(0)}(r\equiv\vert\vec x\vert)\quad .
\end{equation}
For spherical symmetry we have the result
\begin{equation}
-\frac{8\pi}{3}\int r^2 dr \overline\varphi\varphi(r) {A^{(0)}(r)}^2 + 4\pi m^2\int r^2 dr \overline\varphi\varphi\quad ,
\end{equation}
thus, for not too large values of the mass (or for strong background) we have instabilities, or else, with a negative effective mass squared, phanton like equation of state,
\begin{eqnarray}
\rho &=& \overline {\dot\varphi}\dot\varphi - m_{eff}^2 \overline\varphi\varphi + \overline\varphi^\prime\varphi^\prime\quad ,\\
p  &=& \overline {\dot\varphi}\dot\varphi + m_{eff}^2 \overline\varphi\varphi - \overline\varphi^\prime\varphi^\prime \quad ,
\end{eqnarray}
implying the above claim.

We thus found a very simple model where phantom is realized at a local cosmological site, realizing \cite{wanggongabdalla2005}. We just need a solution with $\frac 23 e^2 A(r) > m^2$ at an important neighborhood to have $w< -1$. Small masses, large couplings and nonpreturbative solutions are the requirements and may happen.

Using the Nielsen-Olesen solution this can be arranged providing a local $w< -1$ solution. Indeed, in  \cite{wanggongabdalla2005}, it is shown that at a certain region in redshift space observations point at $w$ crossing -1.

\end{document}